%% file: ANA-HIGP-2024-24-PAPER.tex
\documentclass[PAPER, atlasdraft=false, cernpreprint, USenglish, texmf, orcidlogo]{atlasdoc}
\usepackage{atlaspackage}
\usepackage{atlasbiblatex}

\usepackage{atlasphysics}
\graphicspath{{logos/}{figures/}}

\usepackage{ANA-HIGP-2024-24-PAPER-defs}
\usepackage{multirow}
\usepackage{xspace}
\usepackage{pgffor}

\newcommand{\TBD}{\textcolor{red}{\textbf{TBD}}}
\newcommand{\kappat}{$\kappa_t$}

\AtlasTitle{Probing the Higgs--top Yukawa interaction in the \ttH and \tH processes using \Hyy with the ATLAS detector}

\AtlasAbstract{%
A study of the structure of the coupling between the Higgs boson and the top quark is performed using events from \ttH and \tH production in the \Hyy decay channel, with 164~\ifb of proton--proton collision data at a center-of-mass energy of \com = 13.6~\TeV collected by the ATLAS detector at the LHC\@. The cross section of the \ttH process times the Higgs boson to diphoton decay branching ratio is measured to be $\ttHXSxBRObsTot~\unit{fb}=\ttHXSxBRObs~\unit{fb}$, corresponding to \ttHmuObsTot
times the Standard Model prediction. An observed 95\% confidence level upper limit on the \tH\ production cross section times the Higgs boson to diphoton decay branching ratio is set at 6.2 times the Standard Model prediction, compared to an expected limit of 4.4 times, constituting the most stringent \tH\ upper limit achieved in a single measurement to date. The results are combined with 140~\ifb of proton--proton collision data collected at \com~=~13~\TeV in the same production and decay channel, and a \CP-mixing angle
of $|\alpha| > 38\dg$
is excluded at the 95\% confidence level,
with a purely \CP-odd Higgs-top Yukawa coupling excluded at the level of 5.8 standard deviations, providing the most stringent direct constraints on the \CP structure of the Higgs–top Yukawa interaction to date.
}

\AtlasRefCode{HIGP-2024-24}

\PreprintIdNumber{CERN-EP-2026-150}

\AtlasJournal{Phys.\ Rev.\ Lett.}
\AtlasCoverEgroupAnalysisTeam{atlas-higp-2024-24-analysis-team@cern.ch}

\hypersetup{pdftitle={ATLAS document},pdfauthor={The ATLAS Collaboration}}

\begin{document}

\maketitle

The discovery of the Higgs boson~\cite{HIGG-2012-27,CMS-HIG-12-028} and the observation of most of its production modes~\cite{HIGG-2018-57,HIGG-2018-04,HIGG-2018-13,CMS-HIG-22-001,CMS-HIG-17-035} at the Large Hadron Collider (LHC) offer new opportunities to search for physics beyond the Standard Model (SM) through precise characterization of Higgs boson properties.
In the SM, the Higgs boson is a scalar particle ($J^{CP}$~=~0$^{++}$), where $J$, $C$, and $P$ denote spin, charge conjugation, and parity, respectively~\cite{Gunion:1989we}. The observation of a non-zero \CP-odd component in any Higgs interaction would constitute unambiguous evidence for physics beyond the SM\@.
Although the possibility that the observed Higgs particle is a pure pseudoscalar ($J^{CP}$~=~0$^{+-}$) has been experimentally excluded~\cite{ATLAS:2015zhl,CMS:2021nnc}, a \CP-mixed state remains compatible with current observations. Such a scenario would provide new sources of \CP violation that could help generate the necessary conditions for baryogenesis~\cite{Sakharov:1991,Bahl:2022yrs}. Measurements of the Higgs boson CP structure through Yukawa interactions provide unique sensitivity because fermions can couple directly to pseudoscalar states at tree level, while \CP-odd contributions to Higgs–vector-boson couplings arise only at loop level~\cite{Fontes:2017zfn, Boto:2024jgj}.
Among the Yukawa interactions, the Higgs--top coupling is the largest in the SM\@. It drives the dominant radiative corrections to the Higgs boson mass underlying the hierarchy problem~\cite{Giudice:2008bi}, controls the running of the Higgs quartic coupling that determines electroweak vacuum stability~\cite{Buttazzo:2013uya}, and provides a uniquely sensitive probe of the \CP structure of the Higgs sector~\cite{PhysRevD.104.055045}. Measuring both the strength and the \CP structure of this coupling is therefore of particular importance.

A generic parameterization of the Higgs--top Yukawa interaction can be expressed as~\cite{Demartin:2014fia}
\begin{align*}
\cal{L} = &-\frac{m_t}{v} \left\{ \bar{\psi_{t}}\kappa_t \left[\cos(\alpha) + \text{i}\sin(\alpha)\gamma_{\text{5}} \right]\psi_{t} \right\} H,
\end{align*}
where $m_{t}$ is the top quark mass, $v$ is the Higgs vacuum expectation value, $\kappa_t$~($>0$) is the top Yukawa coupling parameter, \aCP is the \CP-mixing angle, and $\gamma_{\text{5}}$ is a Dirac matrix. The parameter $\kappa_t$ characterizes the overall strength of the coupling with the SM corresponding to $\kappa_t=1$, while \aCP captures its \CP structure, with the SM corresponding to a purely \CP-even coupling at $\alpha=0$. A \CP-odd component appears when $\alpha$ deviates from zero.%

The Higgs--top Yukawa interaction can be probed both directly and indirectly at the LHC\@. It can be constrained indirectly through loop-induced processes such as gluon--gluon fusion (ggF) production and Higgs boson decays into photons~\cite{Brod:2013cka}, assuming no other contributions to these loops~\cite{Ellis:2013yxa}. Measurements of the associated production of the Higgs boson with a top quark pair (\ttH) or with a single top quark (\tH) provide direct access to the Higgs--top Yukawa interaction, including its strength and \CP structure. The destructive interference between \tH production diagrams involving Higgs--top and Higgs--$W$ in the SM is lifted with increasing $\alpha$, enhancing the \tH cross section by a factor of five in the pure \CP-odd scenario compared to a corresponding reduction in the \ttH cross section by a factor of two~\cite{Demartin:2015uha}. The simultaneous measurement of \ttH and \tH therefore offers the strongest direct sensitivity to the Higgs--top coupling \CP structure.

Previously, the ATLAS Collaboration reported constraints on the \CP properties of the Higgs--top Yukawa interaction using measurements of the \ttH, \tHjb, and \tWH production modes in the diphoton~\cite{HIGG-2019-01}, $b$-quark pair~\cite{HIGG-2020-03}, and multilepton final states~\cite{HIGP-2024-08}, as well as measurements of the simultaneous production of four top quarks~\cite{TOPQ-2021-08}, based on the full Run-2 dataset of 140~\ifb of proton--proton collisions at \com = 13~\TeV. Among these channels, the diphoton final state yielded the most stringent constraints, rejecting a pure \CP-odd scenario at 3.9 standard deviations and excluding mixing angles $|\alpha| > 43\dg$ ($63\dg$ expected) at the 95\% confidence level. In addition, the ATLAS Collaboration set constraints on the \tH\ production cross section, with observed (expected) limits of 10 (6.8) and 13.6 (6.1) times the SM prediction in the diphoton channel~\cite{HIGG-2020-16} and in the combination of the $b\bar{b}$, $\tau\tau$, $ZZ^{}$, and $WW^{}$ channels~\cite{HIGP-2024-03}, respectively. The CMS Collaboration also reported similar results on \CP studies and \tH\ searches based on its full Run-2 dataset~\cite{CMS-HIG-19-013,CMS-HIG-21-006,CMS-HIG-19-008,CMS-HIG-19-015,CMS-HIG-19-011}.

This Letter presents a new measurement of the \ttH\ and \tH\ processes in the diphoton channel, optimized simultaneously for the search for \tH\ production and for constraining the \CP-mixing angle. The analysis uses $pp$ collision data at a center-of-mass energy of \com = 13.6~\TeV with an integrated luminosity of 164~\ifb, collected between 2022 and 2024. To maximize sensitivity, this measurement is further combined with the ATLAS measurement of the \CP properties of the Higgs--top Yukawa interaction in the diphoton channel, based on 13~\TeV data collected during LHC Run~2~\cite{HIGG-2019-01}, corresponding to an integrated luminosity of 140~\ifb.

The ATLAS experiment~\cite{PERF-2007-01,GENR-2019-02} at the LHC is a multipurpose particle detector
with a forward--backward symmetric cylindrical geometry and a near \(4\pi\) coverage in
solid angle.\footnote{\AtlasCoordFootnote}
It consists of an inner tracking detector surrounded by a thin superconducting solenoid, electromagnetic and hadronic calorimeters, and a muon spectrometer.
A two-level trigger system was used to select events~\cite{TRIG-2016-01,TRIG-2022-01}. The events used in this analysis were selected by triggers requiring two photons with a $\textit{medium}$ identification requirement~\cite{EGAM-2021-01} and transverse energies of at least 35~\GeV (25~\GeV) for the leading (subleading) photon.

Simulated samples of SM Higgs boson production processes
via \ttH, vector boson fusion (VBF), and associated production with a vector boson (VH)
were generated at next-to-leading order (NLO) in quantum chromodynamics (QCD) with \POWHEGBOX[v2]~\cite{Nason:2004rx,Frixione:2007vw,Alioli:2010xd}, using the \PDFforLHC[21]~\cite{Ball:2022oua} parton distribution function (PDF) set.
The ggF production was simulated with \POWHEGBOX[v2] at next-to-next-to-leading order (NNLO) in QCD using the same PDF set~\cite{Monni:2019whf,Monni:2020nks}.
The \tH process was simulated separately for single-top Higgs production with a forward jet (\tHjb) and single-top Higgs production with a $W$ boson (\tWH), using \MGNLO[3.5.3]~\cite{Alwall:2014hca} at NLO in QCD with the \NNPDF[3.0nlo]~\cite{Ball:2014uwa} PDF set for the matrix element (ME) evaluation.
All samples were interfaced to \PYTHIA[8.310]~\cite{Sjostrand:2014zea} for parton showering (PS) and hadronization using the \NNPDF[2.3lo]~\cite{Ball:2012cx} PDF set.
The five-flavor-scheme (5FS) was used for the \tWH process where the $b$-quark is treated as a massless parton, while the four-flavor-scheme (4FS) was used for \tHjb that assumes the $b$-quark from gluon splitting is massive in the calculation~\cite{PhysRevLett.102.182003,Maltoni:2012pa,Bothmann:2017jfv}. These generator setups were also used to simulate \ttH and \tH \CP samples for different $\alpha$ values, and additional \tH samples with different $\kappa_t$ values within the Higgs Characterization model~\cite{Demartin:2014fia}. The A14 parameter set~\cite{ATL-PHYS-PUB-2014-021}, tuned to data, was used for both PS and the underlying event (UE).

The \CP samples are used to derive parameterization functions of the  $t\bar{t}H$ and $tH$ yields as functions of $\alpha$ and $\kappa_t$, which are used in the statistical interpretations.
Samples interfaced to \HERWIG[7]~\cite{Bahr:2008pv,Bellm:2015jjp} for PS, hadronization, and UE modeling are used for systematic uncertainty studies.
The simulated Higgs boson samples are normalized to the SM cross sections~\cite{LHCHiggsCrossSectionWorkingGroup:2016ypw,Anastasiou:2015ema,Anastasiou:2016cez,Actis:2008ug,Anastasiou:2008tj,Ciccolini:2007jr,Ciccolini:2007ec,Bolzoni:2010xr,Brein:2003wg,Altenkamp:2012sx,Denner:2011id,Beenakker:2002nc,Dawson:2003zu,Yu:2014cka,Frixione:2015zaa,Dawson:2003kb,Dittmaier:2003ej,Harlander:2011aa,Demartin:2015uha} multiplied by the SM branching ratio to diphotons~\cite{LHCHiggsCrossSectionWorkingGroup:2016ypw,Djouadi:1997yw,Djouadi:2006bz,Bredenstein:2006ha,Bredenstein:2006rh} for a Higgs boson mass of 125.09~\GeV~\cite{HIGG-2014-14}.

While the background estimation is primarily data-driven, a simulated \ttyy sample was produced to develop the background model. This sample was generated as a $2\to 8$ body process at leading order with the \MGNLO[3.5.1] generator using the \NNPDF[2.3lo] PDF set and showered with \PYTHIA[8.308].

All generated SM Higgs boson events were passed through a full \GEANT~\cite{Geant4} simulation of the
ATLAS detector~\cite{SOFT-2010-01} and subsequent event reconstruction~\cite{ATLAS:2024vdo}.
The events in the \CP samples with different \aCP and $\kappa_t$
were processed with a fast simulation
in which only the full calorimeter simulation is replaced with a parameterized response~\cite{ATL-PHYS-PUB-2010-013}.
The effects of multiple $pp$ interactions
in the same or neighboring bunch crossings were included.
Events were simulated~\cite{SIMU-2020-01} and weighted to match the distribution of the average number of interactions per bunch
crossing observed in data, which was typically around 60.

Selected events are required to contain two
photon candidates that pass the \textit{loose} isolation criteria~\cite{EGAM-2018-01} and
the \textit{tight} identification criteria~\cite{EGAM-2021-01}. The diphoton invariant mass must satisfy $\myy \in [105,160]$~\GeV, and additional kinematic requirements are imposed on the photons such that the transverse momentum (\pt) of the leading (subleading) photon fulfills $\pt/\myy > 0.35~(0.25)$ and pseudorapidity satisfies $|\eta|<2.37$, excluding the transition region between the barrel and end cap calorimeters, $1.37 < |\eta| < 1.52$. Jets are reconstructed with the anti-\kt algorithm~\cite{Cacciari:2008gp,Fastjet} using a radius parameter $R=0.4$ and are required to have $\pt>20$~\GeV and $|\eta|<4.5$. Identification of jets originating from $b$-hadrons is performed using the transformer-based neural network GN2~\cite{ATLAS:2025dkv}, which is calibrated in a set of exclusive working points corresponding to pseudo-continuous fixed $b$-tagging efficiency intervals of
[0\%, 65\%], [65\%, 70\%], [70\%, 77\%], [77\%, 85\%], [85\%, 90\%] and [90\%, 100\%].
At least one reconstructed $b$-tagged jet satisfying the inclusive 85\% GN2 working point is required in each event.
Isolated muon (electron) candidates with $\pt > 5~(10)$~\GeV\ and $|\eta|<2.7~(2.47)$ that pass \textit{medium} identification requirements~\cite{EGAM-2021-01, MUON-2018-03} are used to separate events in which top quarks decay semileptonically from those that do not.

The analysis is based on a four-stage event classification enabled by cut-based selections and graph neural network (GNN) classifiers.
Reconstructed object kinematics, charge, and pseudo-continuous $b$-tagging scores define the node input features and angular separations define the edge features, following the procedure described in Ref.~\cite{TOPQ-2021-08}.
The GNN model weights up to the output layer are initialized from a classifier trained to discriminate between six Higgs boson production processes and six top quark production processes~\cite{ho2026pretrainedeventclassificationmodel}. This pre-training strategy reduces training convergence times by more than 50\%. Two GNN models are developed using this approach. A multiclass GNN classifier is trained to produce five probability discriminants identifying the three Higgs--top associated production modes, other SM Higgs boson production modes (resonant background), and continuum diphoton production (continuum background). All \CP-even, \CP-mixed, and \CP-odd signal samples are used in this training to maintain discrimination performance under all \CP hypotheses. During training, the continuum background is represented by a data control sample defined by the nominal event selection with at least one photon failing either or both of the \textit{tight} identification and isolation criteria. This definition supplies the GNN with a sufficiently large training dataset that better represents continuum background processes than simulation. An additional binary classification GNN is trained to discriminate between \CP-even-like and \CP-odd-like signal topologies. The \CP classifier is trained using simulation generated with positive $\alpha$ and is therefore insensitive to its sign. A \CP-odd coupling enhances the Higgs boson transverse momentum, the top quark transverse momentum in \tHjb production, and increases (decreases) the top--Higgs angular separation in \ttH and \tWH (\tHjb) production.  Feature ablation studies show that the \CP GNN classification performance is most sensitive to the jet $\pt$ and $\eta$, photon $\eta$ and $\phi$, missing transverse energy and relative azimuthal correlations.

In the first categorization stage, the preselected events with two photons and at least one $b$-tagged jet are filtered into the Hadronic and Leptonic categories, defined by the presence of three or more jets and no isolated leptons, or the presence of at least one isolated lepton, respectively. The distinct event topologies and background composition between the Hadronic and Leptonic categories motivate this separation. In the second stage, events are further categorized into \ttH-, \tWH-, and \tHjb-enriched classes based on the maximum value among the three signal probabilities predicted by the multiclass GNN\@.
This categorization step reduces the correlation between the \ttH and \tH rate measurements. In the third stage, events in each signal-enriched class are assigned to one of three categories ranging from more \CP-even-like to more \CP-odd-like based on boundaries of the binary \CP GNN discriminant. This categorization provides sensitivity to rate and kinematic shape effects induced by varying the \CP-mixing angle. In the fourth stage, each \CP category is subdivided based on boundaries of the averaged continuum and resonant background probabilities predicted by the multiclass GNN\@. This final splitting step is performed to maximize sensitivity to the \ttH and \tH signal processes separately. This four-stage event categorization strategy allows the measurement to be sensitive to both the \CP properties of the Higgs–top Yukawa interaction and the production cross sections of multiple Higgs–top associated production modes.

In total, 32 categories are defined. The boundaries for each of the \CP and background discriminants used to define the final event categorization are simultaneously optimized to maximize the expected exclusion of the maximally \CP-mixing scenario and expected significance of the SM \tH signal process based on counting significance. Additional details of the categorization are provided in the End Matter. To illustrate the contribution from the most sensitive categories, the diphoton invariant mass distributions of the 32 categories are summed with weights of $\ln(1+S/B)$, where $S/B$ is the expected signal-to-background ratio in the corresponding category, as shown in Figure~\ref{fig:weightedmass}.

\begin{figure}[h]
\centering
\includegraphics[width=0.5\linewidth]{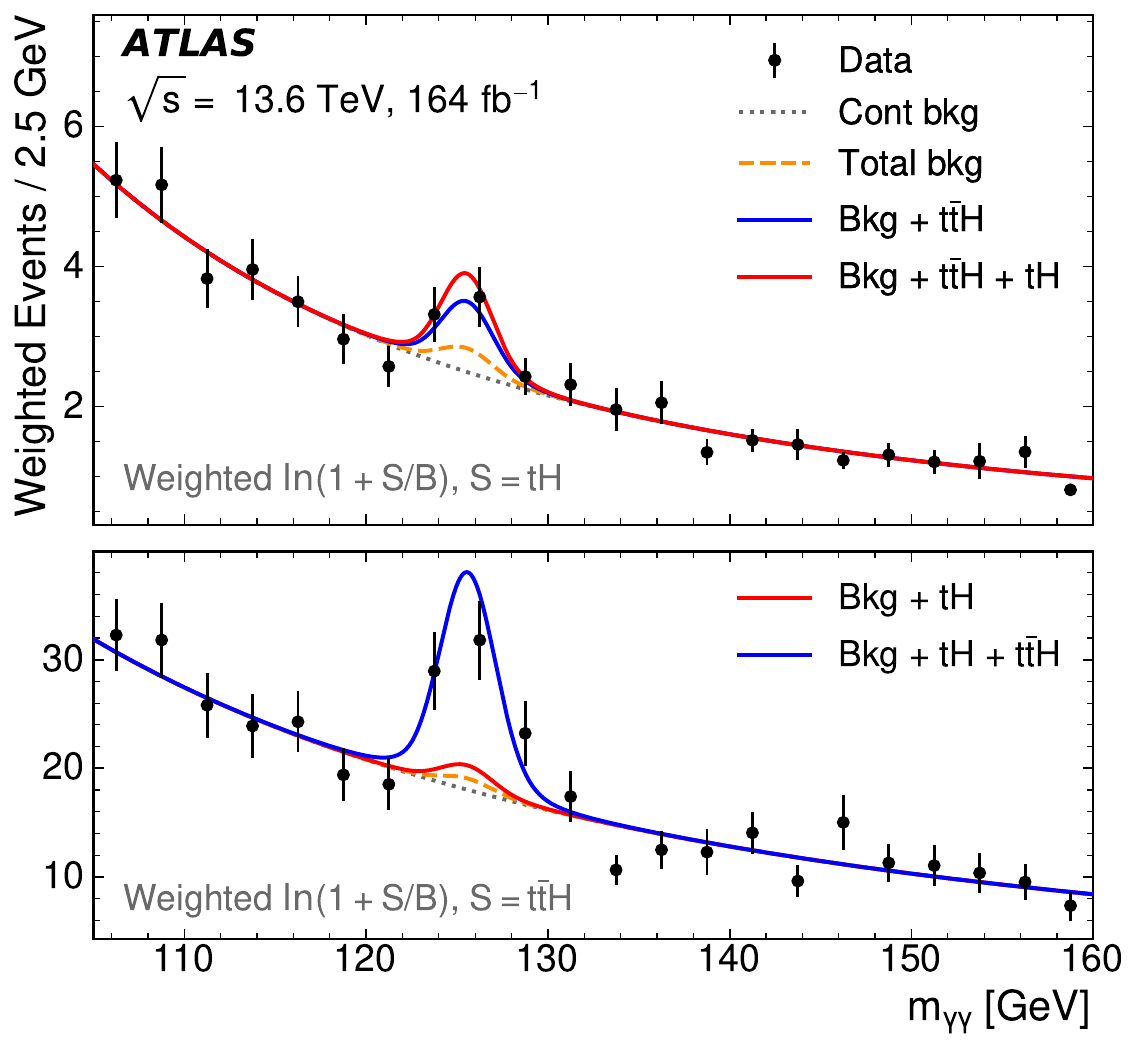}
\caption{Diphoton invariant mass distribution (\myy) obtained by summing the 32 event categories, each weighted by $\ln(1+S/B)$, where $S/B$ denotes the expected signal-to-background ratio in that category. The signal is defined as the \tH (\ttH) production, while the background includes the \ttH (\tH) production mode and resonant and continuum background processes in the top (bottom) panel.
The mass window used to compute the weight is defined as the smallest interval containing 90\% of the signal yield.
The fitted stacked contributions of the continuum background (gray), resonant background (orange), \ttH (blue), and \tH (red) processes are shown compared to the data.
}
\label{fig:weightedmass}
\end{figure}

A likelihood function is constructed from the data and the probability density functions of the diphoton invariant mass distributions in the 32 event categories. The parameterization of the signal and background invariant mass distributions follows the procedure outlined in Ref.~\cite{HIGG-2020-16}. The continuum background is modeled by either a first-order power law or an exponential function, chosen per category to best describe the \ttyy diphoton invariant mass spectrum, and fitted to data. The resonant Higgs boson backgrounds are modeled by double-sided Crystal Ball functions, derived from simulation and normalized to the SM prediction. Measurements of the SM \ttH and \tH production rates are obtained by fitting the corresponding signal strength parameters to data, while the \CP study is performed by parameterizing the \ttH and \tH production rates as a function of \kappat~and $\alpha$. The per-category event yields are shown in the End Matter. Systematic uncertainties are incorporated into the likelihood function as nuisance parameters with associated Gaussian constraint terms. The profile likelihood ratio is used as the test statistic, and the asymptotic approximation~\cite{Cowan:2010js} is used for statistical interpretation.

The total measurement uncertainty is dominated by the statistical uncertainty of the data, with subleading contributions associated with the continuum background model, theoretical modeling, and experimental systematic uncertainties.
An uncertainty is defined to account for possible mis-modeling of the continuum background in the diphoton invariant mass range \myy $\in [123, 127]$~\GeV following the strategy of previous measurements~\cite{HIGG-2020-16}.
Theoretical uncertainties on the \ttH and \tH rates are evaluated in each category. Uncertainties of the \ttH and \tH cross sections~\cite{LHCHiggsCrossSectionWorkingGroup}, \Hyy branching fraction~\cite{LHCHiggsCrossSectionWorkingGroup:2016ypw}, and Higgs boson mass~\cite{ATLAS:2015yey} are included. QCD scale uncertainties are derived from the envelope of variations due to the choice of renormalization and factorization scales and account for missing higher-order terms in perturbative QCD calculations. Uncertainties for PS, hadronization, and UE are evaluated with alternative generators and uncertainties in the PDF and the value of the strong coupling constant $\alpha_s$ are determined from the \PDFforLHC[21] eigenvector and $\alpha_s$ variations.
Conservative uncertainties of 100\% are applied to the ggF, VBF, and $VH$ rates arising from imprecise modeling of associated heavy-flavor jet production.
Experimental uncertainties arise from the photon trigger efficiency and the identification criteria, isolation criteria, energy scale, and resolution for photons, electrons, and muons~\cite{ATLAS:2019dpa, EGAM-2018-01, PERF-2015-10}.
Jets have uncertainties from $b$-tagging~\cite{FTAG-2018-01}
and vertex identification~\cite{PERF-2014-03}
in addition to their energy scale and resolution~\cite{JETM-2018-05}. Uncertainties in the measurement of
$E_\text{T}^\text{miss}$~\cite{JETM-2020-03},
which is used primarily in the Leptonic categories, are also included.
Furthermore, the integrated luminosity uncertainty for the Run-3 dataset is 1.9\%, determined using LUCID-2~\cite{LUCID2}.

Assuming a purely \CP-even coupling, the observed (expected) significance for rejecting the background-only hypothesis is \ttHSigObs (\ttHSigExp) for the \ttH process and \tHSigObs (\tHSigExp) for the \tH process.
The signal strength of \ttH\ is observed to be $\ttHmuObsTot=\ttHmuObs$~times the SM expectation. This corresponds to a \ttH cross section times the \Hyy branching ratio at \com = 13.6~\TeV of $\ttHXSxBR = \ttHXSxBRObsTot~\unit{fb}=\ttHXSxBRObs~\unit{fb}$, obtained with removing theoretical uncertainties of the \ttH/\tH cross section and \Hyy branching ratio. The signal strength of \tH is measured to be $\tHmuObsTot=\tHmuObs$. A 95\% confidence level (CL) upper limit on its cross section is set using the $\mathrm{CL}_{\mathrm{s}}$ method~\cite{Read:2002hq}, with the observed and expected limits found to be \tHLimitObs and \tHLimitExp times the SM prediction, respectively. In these measurements, the rates of all non-\ttH/\tH\ processes are fixed to their SM-predicted values, while the rates of the \CP-even \ttH\ and \tH\ processes are left unconstrained.
The expected \tH sensitivity in this measurement is improved by approximately $30\%$ relative to the Run-2 measurement in the \Hyy\ decay channel~\cite{HIGG-2020-16}, driven predominantly by the updated analysis strategy and by improved object reconstruction and identification, in particular the GN2 $b$-tagging algorithm~\cite{ATLAS:2025dkv}.
As illustrated in Figure~\ref{fig:coupling_strength}, this new strategy also leads to a significantly reduced correlation between the \ttH\ and \tH\ production measurements, from $-44\%$~\cite{HIGG-2020-16} to $-28\%$, enabled by dedicated \tH-enriched categories that isolate \tH-like from \ttH-like event topologies and thereby reduce the
degeneracy between their contributions in the fit.

The theoretical and experimental systematic uncertainties on \ttHXSxBR are 4\% and 12\%, respectively, compared to a statistical uncertainty of 22\%. The sum of jet energy scale and resolution uncertainties has the largest single impact, at the 9\% level, and the photon trigger, identification and isolation efficiency uncertainties total to 4\%.

\begin{figure}[htpb]
\centering
\includegraphics[width=0.45\textwidth]{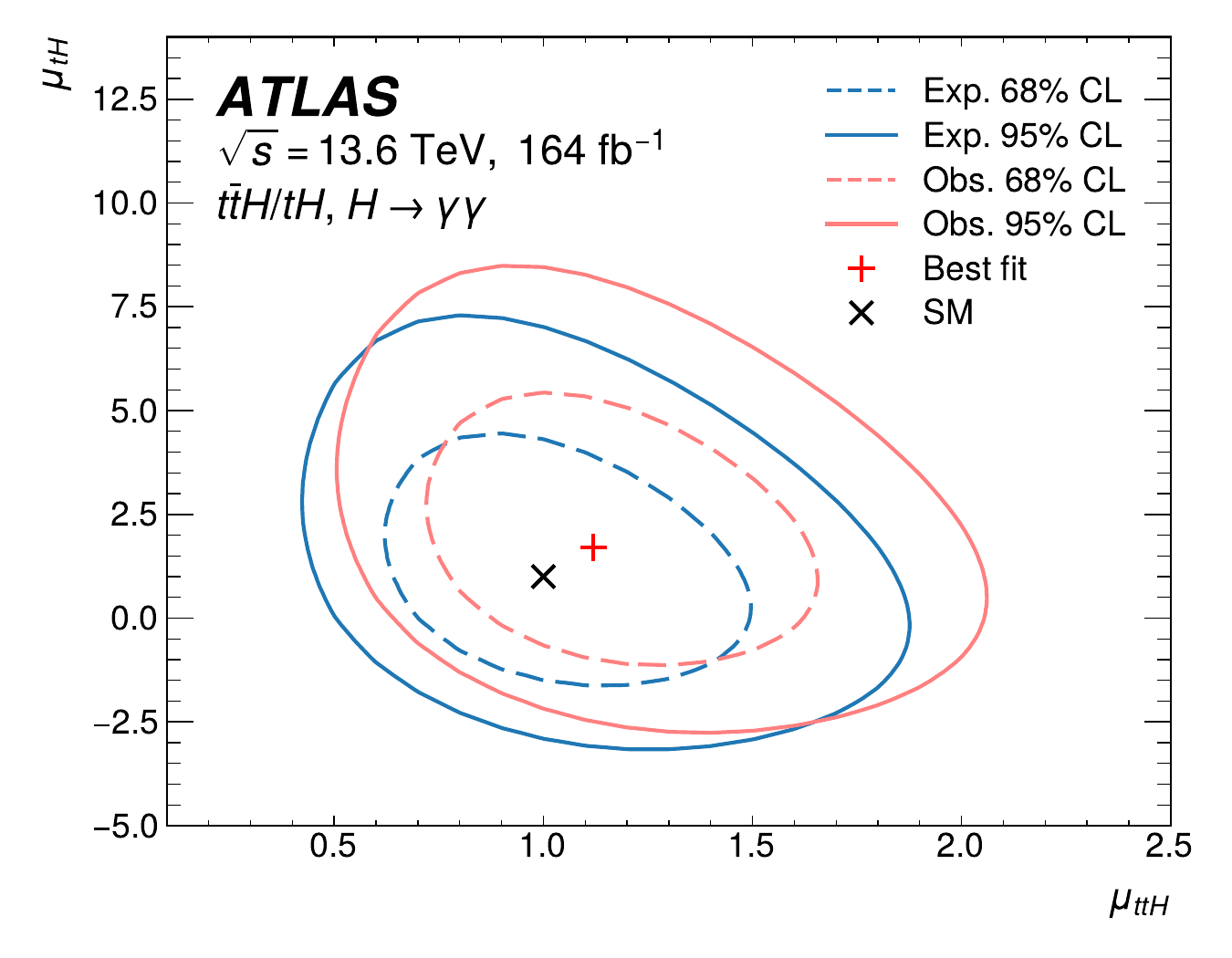}
\caption{Two-dimensional likelihood contours for the \tH signal strength ($\mu_{\tH}$) and the \ttH signal strength ($\mu_{\ttH}$) obtained from the Run-3 measurement.}
\label{fig:coupling_strength}
\end{figure}

The tree-level extraction of the Higgs--top Yukawa coupling requires additional information; specifically, the rate of the Higgs boson decay into diphotons and the rate of the small ggF background contribution must be constrained using external measurements. A procedure identical to that of Ref.~\cite{HIGG-2019-01} is adopted: the effective Higgs boson coupling modifiers for the Higgs boson decay into diphotons, $\kappa_\gamma$, and for ggF, $\kappa_g$, are constrained using measurements reported in Ref.~\cite{HIGG-2021-23}, which includes Run-2 ATLAS Higgs boson measurements, including \ttH\ and \tH\ measurements in the diphoton and other decay channels. To avoid double counting
and to ensure a consistent interpretation of the Higgs--top Yukawa coupling structure,
the categories enriched in \ttH\ and \tH production are removed from the Run-2 measurements, and a two-dimensional Gaussian constraint term for $\kappa_g$ and $\kappa_\gamma$ is extracted and applied for the Higgs--top \CP interpretation. The impact of removing these categories has a negligible effect on the derived constraint. The correlations of systematic uncertainties between this measurement and the remaining inputs in Ref.~\cite{HIGG-2021-23} are neglected. The single effective coupling modifier $\kappa_g$ does not capture \CP-induced modifications of the ggF kinematics. The resulting per-category rate dependence was verified to be small and is therefore neglected. Modifications of the \Hyy decay rate are accurately parameterized by $\kappa_\gamma$ due to the absence of kinematic effects in two-body decays.

Figure~\ref{fig:CPresult} shows the constraints on the \CP-mixing angle \aCP as exclusion contours in the $\kappa_t\cos(\aCP)$--$\kappa_t\sin(\aCP)$ plane. The parameter $\kappa_t$ is measured to be \ktObs with \aCP fixed at 0. Pseudo-experiments using the Neyman construction~\cite{Neyman:1933} are performed to derive the 95\% CL interval for \aCP,
where event yields are sampled from the probability density function with nuisance parameters fixed to their best-fit values from data,
and values of $|\alpha| > \alphaProfiledLimitObs$ are excluded without prior assumptions on $\kappa_t$. The expected exclusion, evaluated with respect to the pure \CP-even hypothesis, is $|\alpha| > \alphaProfiledLimitExp$ at 95\% CL\@. The observed (expected) significance for rejecting \aCP = 45\dg\ and 90\dg\ is \CPmixExclObs (\CPmixExclExp) and \CPoddExclObs (\CPoddExclExp), respectively.
The sensitivity to the \CP structure is driven by both rate and kinematic effects. While the analysis strategy is not explicitly optimized to target shape effects, the GNN-based categorization provides differential discrimination of \CP-dependent kinematic features, such that the yields across categories encode a coherent pattern sensitive to both normalization and shape contributions.
Compared to the Run-2 \CP analysis~\cite{HIGG-2019-01}, the exclusion significance improves by 55\% for both \aCP = 45\dg\ and 90\dg. Most of the improvement arises from the updated analysis strategy, while the increased size of the data sample due to the higher luminosity and center-of-mass energy contributes 18\% and 22\% for $\aCP = 45\dg$ and $90\dg$, respectively.

\begin{figure}[htbp]
\centering
{\includegraphics[width=0.5\textwidth]{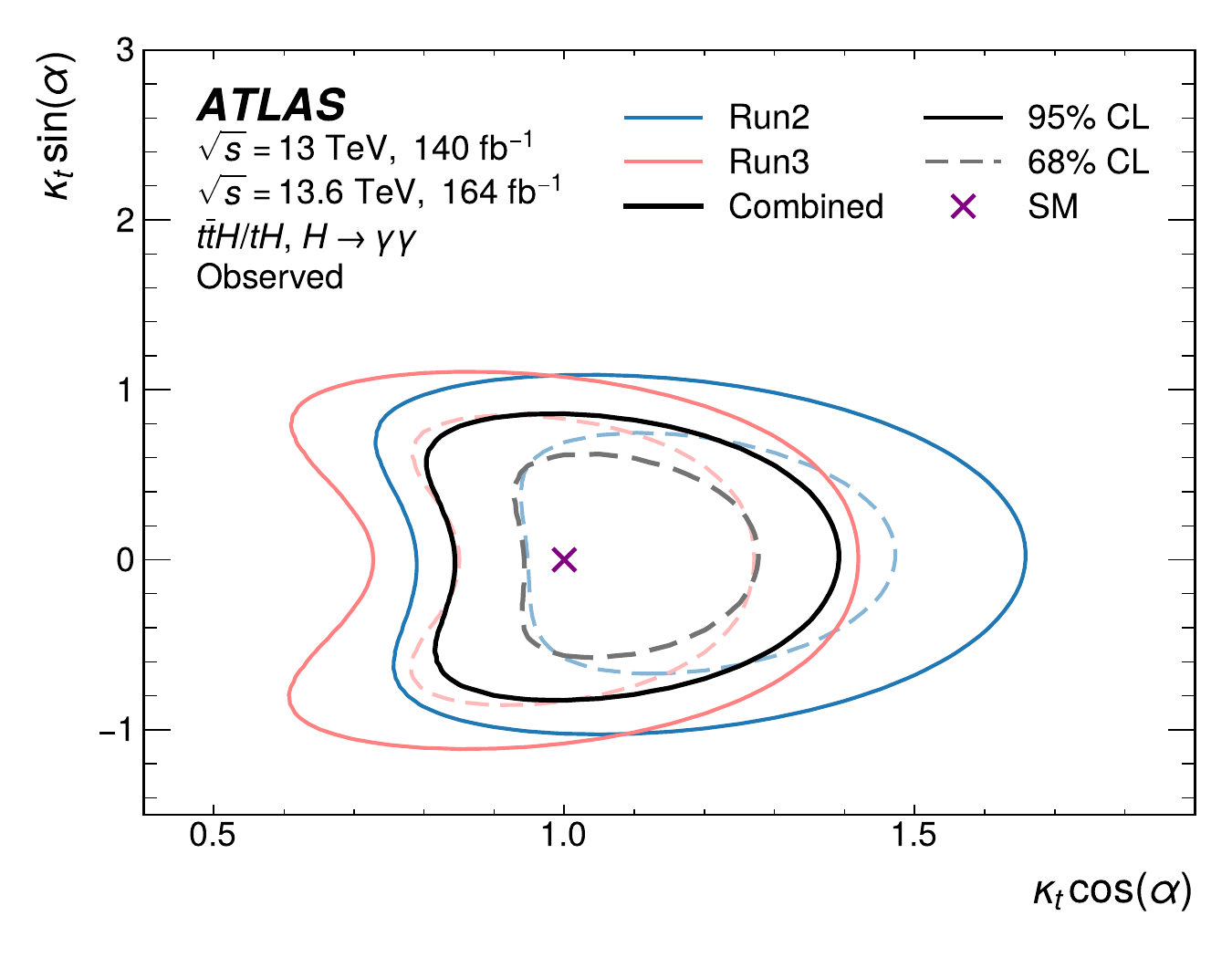}}
\caption{Two-dimensional likelihood contours for $\kappa_t\cos(\alpha)$ and $\kappa_t\sin(\alpha)$ with ggF and $H\to\gamma\gamma$ constrained by the Higgs boson coupling combination. Results from Run-2, Run-3, and the combined measurement are overlaid.}
\label{fig:CPresult}
\end{figure}

To maximize sensitivity to the \CP structure, the likelihood function of the Run-3 measurement is combined with that of the Run-2 measurement~\cite{HIGG-2019-01}, with parameters of interest (e.g., \kappat, $\alpha$, and signal strengths) correlated across datasets, as are nuisance parameters of shared origin, such as those related to theoretical and experimental uncertainties.
The combined Run-2 and Run-3 measurement yields the most stringent constraint on the \CP-mixing angle $\alpha$. The observed (expected) 95\% CL exclusion is found to be $|\alpha| > \CombalphaLimitObs$ (\CombalphaLimitExp) degrees, while the $45\dg$ and $90\dg$ scenarios are rejected with a significance of \CombCPmixExclObs (\CombCPmixExclExp) and \CombCPoddExclObs (\CombCPoddExclExp), respectively.
Assuming a \CP-even coupling, $\kappa_t$ is measured to be \CombktObs.

In summary, a measurement of the Higgs--top Yukawa interaction is performed in the diphoton final state using 164~\ifb\ of data collected during LHC Run~3. The signal strength of the \ttH\ process is found to be $\ttHmuObsTot = \ttHmuObs$, and the corresponding production cross section times the \Hyy branching ratio at \com = 13.6~\TeV is measured to be
$\ttHXSxBR = \ttHXSxBRObsTot~\unit{fb} = \ttHXSxBRObs~\unit{fb}$.
The production rate of the \tH\ process is constrained to be less than \tHLimitObs times the SM prediction at the 95\% CL, providing the most stringent \tH\ upper limit from a single measurement to date. Using this Run-3 dataset, values of the \CP-mixing angle $|\alpha| > \alphaProfiledLimitObs$ are excluded at 95\% CL, while the observed \kappat\ is found to be \ktObs. A statistical combination with the Run-2 measurement~\cite{HIGG-2019-01}
strengthens this constraint to $|\alpha| > \CombalphaLimitObs$ at 95\% CL and yields a $\kappa_t$ of \CombktObs, providing the strongest direct constraint of \CP violation in the Higgs–top interaction to date.

\section*{Acknowledgements}
%

%

%
%

%
%

We thank CERN for the very successful operation of the LHC and its injectors, as well as the support staff at
CERN and at our institutions worldwide without whom ATLAS could not be operated efficiently.

The crucial computing support from all WLCG partners is acknowledged gratefully, in particular from CERN, the ATLAS Tier-1 facilities at TRIUMF/SFU (Canada), NDGF (Denmark, Norway, Sweden), CC-IN2P3 (France), KIT/GridKA (Germany), INFN-CNAF (Italy), NL-T1 (Netherlands), PIC (Spain), RAL (UK) and BNL (USA), the Tier-2 facilities worldwide and large non-WLCG resource providers. Major contributors of computing resources are listed in Ref.~\cite{ATL-SOFT-PUB-2026-001}.

We gratefully acknowledge the support of ANPCyT, Argentina; YerPhI, Armenia; ARC, Australia; BMWFW and FWF, Austria; ANAS, Azerbaijan; CNPq and FAPESP, Brazil; NSERC, NRC and CFI, Canada; CERN; ANID, Chile; CAS, MOST and NSFC, China; Minciencias, Colombia; MEYS CR, Czech Republic; DNRF and DNSRC, Denmark; IN2P3-CNRS and CEA-DRF/IRFU, France; SRNSFG, Georgia; BMFTR, HGF and MPG, Germany; GSRI, Greece; RGC and Hong Kong SAR, China; ICHEP and Academy of Sciences and Humanities, Israel; INFN, Italy; MEXT and JSPS, Japan; CNRST, Morocco; NWO, Netherlands; RCN, Norway; MNiSW, Poland; FCT, Portugal; MNE/IFA, Romania; MSTDI, Serbia; MSSR, Slovakia; ARIS and MVZI, Slovenia; DSI/NRF, South Africa; MICIU/AEI, Spain; SRC and Wallenberg Foundation, Sweden; SERI, SNSF and Cantons of Bern and Geneva, Switzerland; NSTC, Taipei; TENMAK, T\"urkiye; STFC/UKRI, United Kingdom; DOE and NSF, United States of America.

Individual groups and members have received support from BCKDF, CANARIE, CRC and DRAC, Canada; CERN-CZ, FORTE and PRIMUS, Czech Republic; COST, ERC, ERDF, Horizon 2020 and Marie Sk{\l}odowska-Curie Actions, European Union; Investissements d'Avenir Labex, Investissements d'Avenir Idex and ANR, France; DFG and AvH Foundation, Germany; Herakleitos, Thales and Aristeia programmes co-financed by EU-ESF and the Greek NSRF, Greece; BSF-NSF and MINERVA, Israel; NCN and NAWA, Poland; La Caixa Banking Foundation, CERCA and AGAUR programs from Generalitat de Catalunya and PROMETEO and GenT Programmes Generalitat Valenciana, Spain; G\"{o}ran Gustafssons Stiftelse, Sweden; The Royal Society and Leverhulme Trust, United Kingdom; Eric and Wendy Schmidt Fund for Strategic Innovation, United States of America.

In addition, individual members wish to acknowledge support from Chile: Agencia Nacional de Investigaci\'on y Desarrollo (ANID FONDECYT reg. 1230987, FONDECYT 1230812, FONDECYT 1240864, Fondecyt 3240661, Fondecyt Regular 1240721); China: Chinese Ministry of Science and Technology (MOST-2023YFA1605700, MOST-2023YFA1609300), National Natural Science Foundation of China (NSFC 12275265, NSFC-W2543005); Czech Republic: Czech Science Foundation (GACR - 24-11373S), Ministry of Education Youth and Sports (ERC-CZ-LL2327, FORTE CZ.02.01.01/00/22\_008/0004632); EU: H2020 European Research Council (ERC - 101002463); European Union: European Research Council (BARD No. 101116429, ERC - 101219398, ERC 101089007), European Regional Development Fund (HE COFUND GA No.101081355, ERDF), Marie Sklodowska-Curie Actions (GAP-101168829); France: Agence Nationale de la Recherche (ANR-21-CE31-0013, ANR-22-EDIR-0002, ANR-24-CE31-0504-01); Germany: Deutsche Forschungsgemeinschaft (DFG - 469666862); China: Research Grants Council (GRF); Italy: Istituto Nazionale di Fisica Nucleare (LHC-MIUR - 28003/2025), Ministero dell'Università e della Ricerca (NextGenEU I53D23000820006 M4C2.1.1, SOE2024\_0000023); Japan: Japan Society for the Promotion of Science (JSPS KAKENHI  JP25H0063, JSPS KAKENHI JP22H04944, JSPS KAKENHI JP24K23939, JSPS KAKENHI JP24KK0251, JSPS KAKENHI JP25H00650, JSPS KAKENHI JP25H01291, JSPS KAKENHI JP25K01011, JSPS KAKENHI JP25K01023, JSPS KAKENHI JP25KK0047); Poland: Polish National Science Centre (NCN 2021/42/E/ST2/00350, NCN OPUS 2023/51/B/ST2/02507, NCN OPUS nr 2022/47/B/ST2/03059, NCN UMO-2019/34/E/ST2/00393, UMO-2022/47/O/ST2/00148, UMO-2023/49/B/ST2/04085, UMO-2023/51/B/ST2/00920, UMO-2024/53/N/ST2/00869); Spain: Agència de Gestió d'Ajuts Universitaris i de Recerca. (AGAUR - 2023 BP 00141), Ministry of Science and Innovation (RYC2019-028510-I, RYC2020-030254-I, RYC2021-031273-I, RYC2022-038164-I), Ministerio de Ciencia, Innovación y Universidades/Agencia Estatal de Investigaci\'on (EU NextGenerationEU (PRTR-C17.I1), PID2022-142604OB-C22); Sweden: Carl Trygger Foundation (Carl Trygger Foundation CTS 22:2312), Swedish Research Council (Swedish Research Council 2023-04654, VR 2021-03651, VR 2022-03845, VR 2022-04683, VR 2023-03403, VR 2024-05451, VR 2025-05940), Knut and Alice Wallenberg Foundation (KAW 2023.0366); United Kingdom: The Binks Trust, Royal Society (NIF-R1-231091); United States of America: U.S. Department of Energy (ECA DE-AC02-76SF00515), John Templeton Foundation (John Templeton Foundation 63206), Neubauer Family Foundation.

%
%


%
\clearpage
\appendix
\part*{End matter}
\addcontentsline{toc}{part}{Appendix}
Table~\ref{tab:GNNinputs} lists the node features used in the training of all GNN discriminants. Figure~\ref{fig:category_diagram} summarizes the event category definitions based on intervals of the various GNN discriminants obtained from the multiclass GNN and binary \CP GNN\@. Figure~\ref{fig:category_summary} shows the post-fit event yields in the smallest \myy interval containing 90\% of the signal. As expected, the \ttH Yukawa categories have the highest acceptance of \ttH events, while the \tHjb Yukawa categories have the highest acceptance of \tH events. Comparison with the expected total signal yield in the pure \CP-odd case shows that the expected \ttH yield decreases substantially, while signal yields in the \tHjb Yukawa categories increase.

Figure~\ref{fig:category_cp_summary} summarizes the total signal yields in categories optimized to have high acceptance of the \CP-even, \CP-mixed, and \CP-odd signal processes. As expected, the \CP-even and \CP-odd category signal yields are maximized in the \CP-even and \CP-odd scenarios, respectively. Figure~\ref{fig:alpha_CP_1D_toys_obs} shows the observed likelihood scans as a function of \aCP for the Run-2 and Run-3 measurements along with their combination, and the corresponding 95\% confidence level critical line obtained through sampling pseudo-experiments using the Neyman construction~\cite{Neyman:1933}. This calculation is required to obtain confidence intervals with correct coverage because the likelihood function is non-quadratic in the parameter of interest $\alpha$, violating the regularity conditions of the Wald approximation~\cite{Wald1943TestsOS} that underlies the asymptotic formulae of Ref.~\cite{Cowan:2010js}.

\begin{table}[h]
\centering
\caption{\label{tab:GNNinputs} Definitions of the node features used in the GNN training. Photon transverse momentum is divided by the diphoton invariant mass to avoid correlations between the GNN output discriminants and the diphoton invariant mass. Jet flavor information is represented by the pseudo-continuous $b$-tagging (PCBT) scores. The node type is an integer value encoding the type of physics object represented by the node. Features with dimensionful quantities are in units of \GeV and are further scaled by 0.1.}
\begin{tabular}{ l c  c c c c }
\toprule
Feature & Jet & $\mu$ & $e$ & $\gamma$ & $E_\text{T}^{\text{miss}}$ \\
\midrule
\multirow{4}{*}{Kinematic}
& $\pt$ & $\pt$ & $\pt$ & $\pt/\myy$ & $\pt$ \\
& $\eta$ & $\eta$& $\eta$ & $\eta$ & 0 \\
& $\phi$ & $\phi$ & $\phi$ & $\phi$ & $\phi$ \\
& $\pt \times \cosh\eta$ & $\pt \times \cosh\eta$ & $\pt \times \cosh\eta$ & $\pt/\myy \times \cosh\eta$ & \pt\\
\midrule
$b$-tagging & PCBT bin & 0 & 0 & 0 & 0 \\
Charge & 0 & $\pm 1$ & $\pm 1$ & 0 & 0 \\
Node Type & 0 & 1 & 2 & 3 & 4 \\
\bottomrule
\end{tabular}
\end{table}

\begin{figure}[htbp]
\centering
\includegraphics[width=0.5\textwidth]{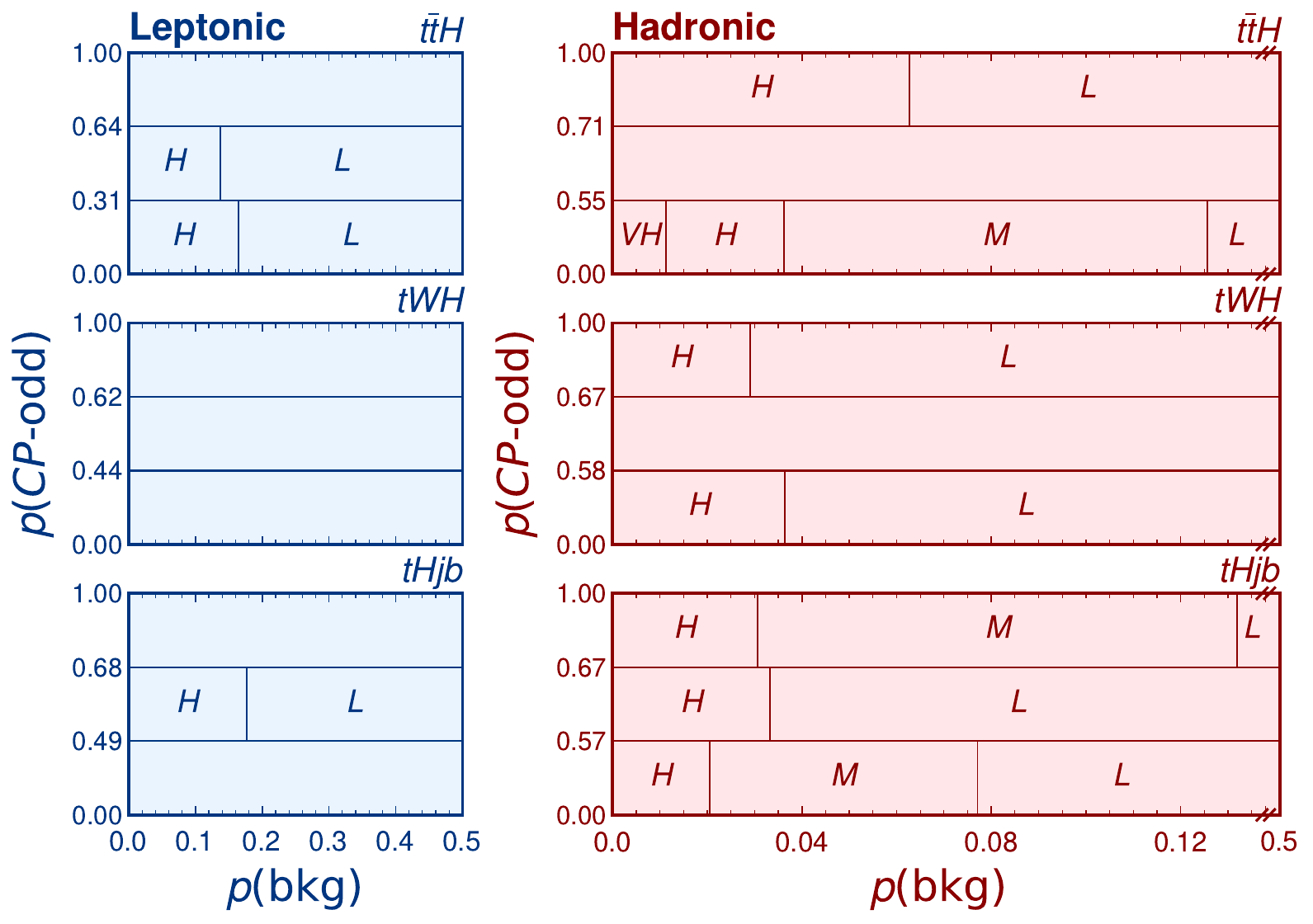}
\caption{A summary of the event category definitions in the Leptonic and Hadronic preselection regions. The \ttH, \tWH, and \tHjb Yukawa categories consist of events that maximize the \ttH, \tWH, and \tHjb GNN discriminants, respectively. In each Yukawa category, three intervals of the \CP GNN discriminant, $p$(\CP-odd), are defined targeting \CP-even, \CP-mixed, and \CP-odd production. Depending on the expected number of events, these categories are further subdivided in intervals of the background GNN discriminant, $p$(bkg), to obtain Low, Medium, High, or Very High signal purity categories (indicated by labels L, M, H, and VH, respectively).}
\label{fig:category_diagram}
\end{figure}

\begin{figure}[htbp]
\centering
\includegraphics[width=0.5\textwidth]{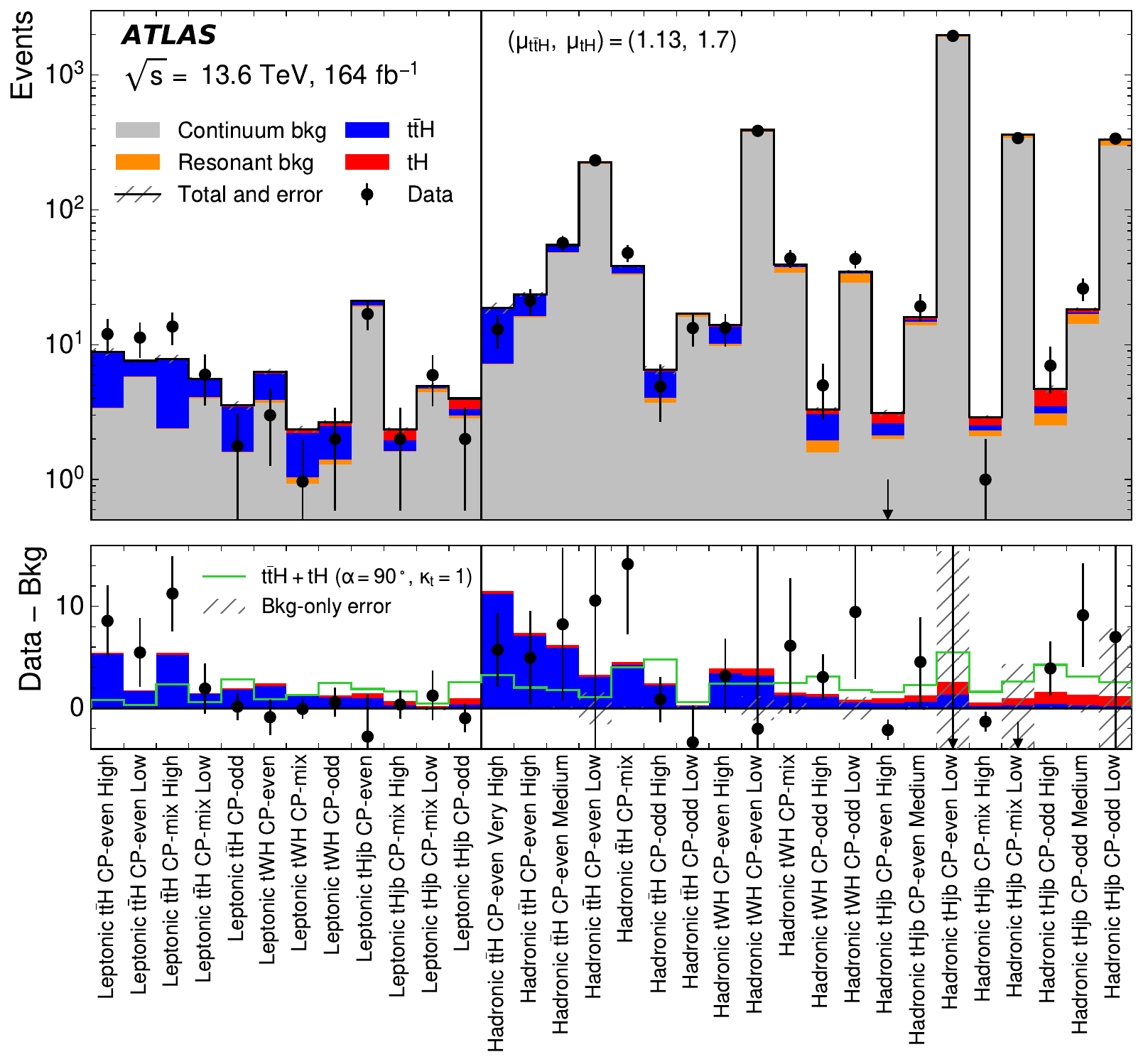}
\caption{Observed data event yields compared to stacked signal and background event yields and their total systematic uncertainty from the result of the simultaneous (\muttH, \mutH) fit. Event yields are calculated within the narrowest diphoton invariant mass interval containing 90\% of the signal yield. The lower panel shows the background-subtracted data yields compared to the background-only uncertainty overlaid with the expected total signal yield in the pure \CP-odd scenario. Arrows indicate the direction of the (background-subtracted) data yield when it lies outside the plot range.}
\label{fig:category_summary}
\end{figure}

\begin{figure}[htbp]
\centering
\includegraphics[width=0.5\textwidth]{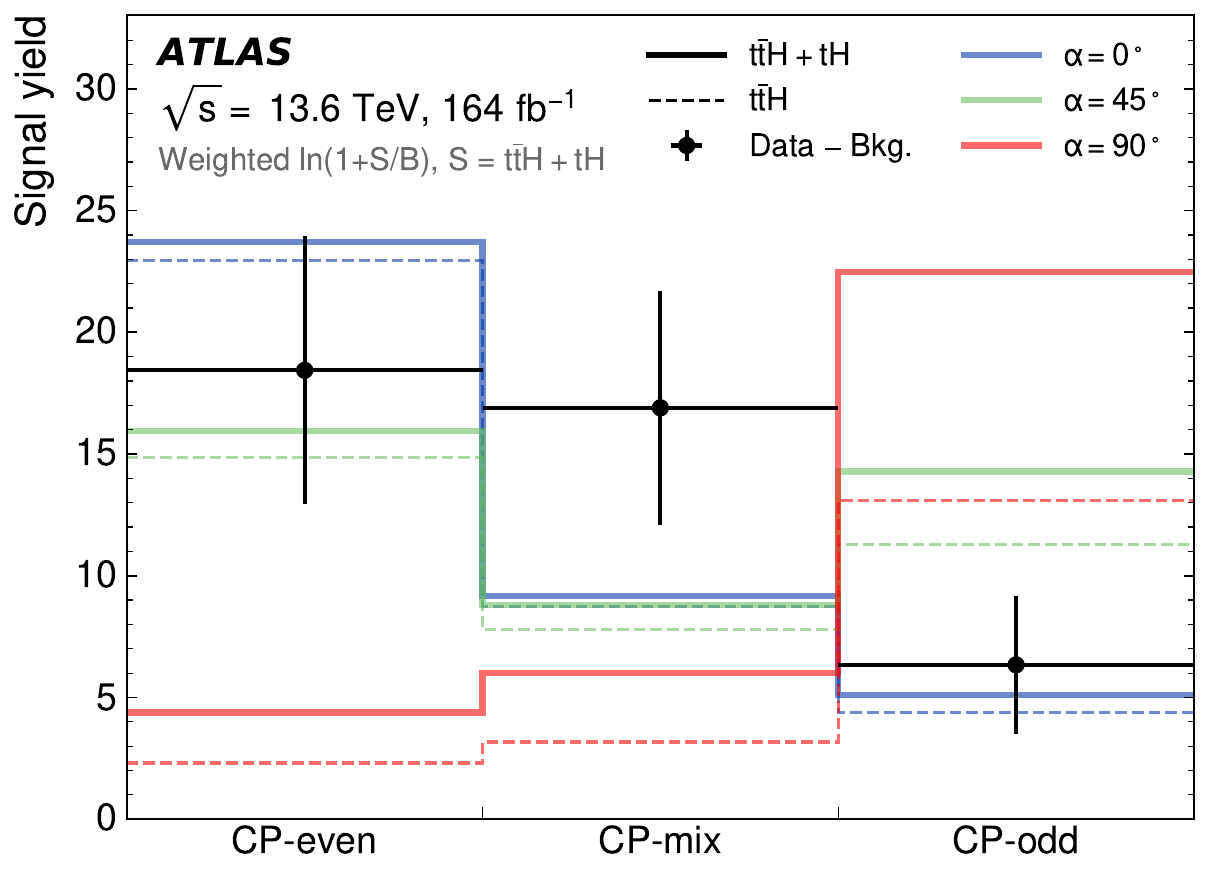}
\caption{The sum of background subtracted data and expected total signal yield in each of the \CP-even, \CP-mixed, and \CP-odd categories,
calculated in the narrowest diphoton invariant mass interval containing 90\% of the signal yield.
The yields are weighted by $\ln(1+S/B)$, where $S/B$ denotes the expected signal-to-background ratio in that category. The signal is defined as the \ttH and \tH production, while the background includes the resonant and continuum background processes.
The mass window used to compute the weight is defined as the interval containing the central 90\% of the signal yield.
}
\label{fig:category_cp_summary}
\end{figure}

\begin{figure}[htbp]
\centering
\includegraphics[width=0.5\textwidth]{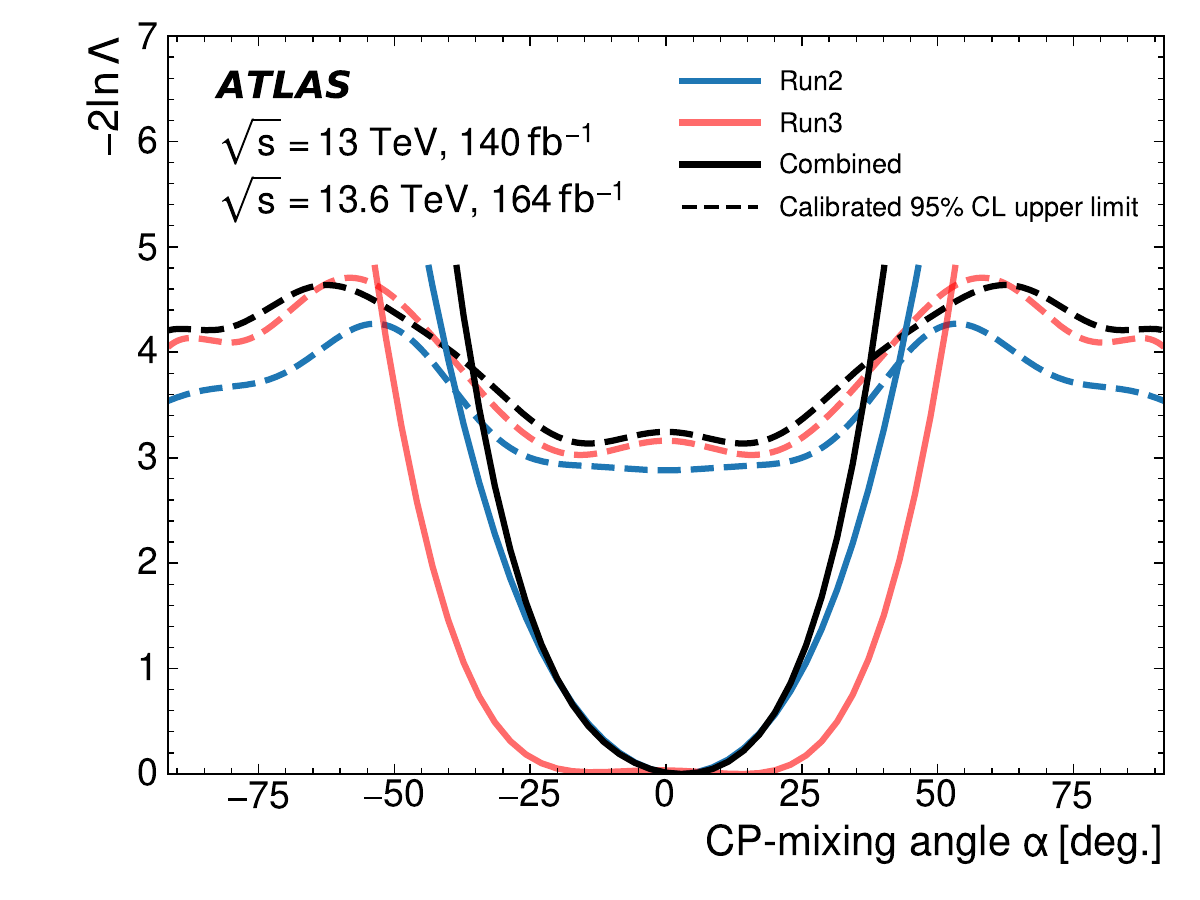}
\caption{Observed likelihood contours for the Run-2, Run-3, and combined measurement of \aCP with the value of $\kappa_t$ profiled to its best fit value. The critical line defining the 95\% quantile of the test statistic's distribution is shown as a dotted line for each measurement and is obtained by sampling pseudo-experiments generated from the observed data.}
\label{fig:alpha_CP_1D_toys_obs}
\end{figure}

\clearpage
\printbibliography
\clearpage
\input{atlas_authlist}
%

%
%

%

%
%
%
%

\end{document}

%% file: atlas_authlist.tex
 
\begin{flushleft}
\hypersetup{urlcolor=black}
{\Large The ATLAS Collaboration}

\bigskip

\AtlasOrcid[0000-0002-6665-4934]{G.~Aad}$^\textrm{\scriptsize 102}$,
\AtlasOrcid[0000-0001-7616-1554]{E.~Aakvaag}$^\textrm{\scriptsize 17}$,
\AtlasOrcid[0000-0002-5888-2734]{B.~Abbott}$^\textrm{\scriptsize 121}$,
\AtlasOrcid[0000-0002-0287-5869]{S.~Abdelhameed}$^\textrm{\scriptsize 83b}$,
\AtlasOrcid[0000-0002-1002-1652]{K.~Abeling}$^\textrm{\scriptsize 54}$,
\AtlasOrcid[0000-0001-5763-2760]{N.J.~Abicht}$^\textrm{\scriptsize 48}$,
\AtlasOrcid[0000-0002-8496-9294]{S.H.~Abidi}$^\textrm{\scriptsize 30}$,
\AtlasOrcid[0009-0003-6578-220X]{M.~Aboelela}$^\textrm{\scriptsize 44}$,
\AtlasOrcid[0000-0002-9987-2292]{A.~Aboulhorma}$^\textrm{\scriptsize 36e}$,
\AtlasOrcid[0000-0001-5329-6640]{H.~Abramowicz}$^\textrm{\scriptsize 154}$
% ...
\bigskip
\\

% ...
$^{17}$Department for Physics and Technology, University of Bergen, Bergen; Norway.\\
% ...
$^{30}$Physics Department, Brookhaven National Laboratory, Upton NY; United States of America.\\
% ...
$^{36}$$^{(a)}$Facult\'e des Sciences Ain Chock, Universit\'e Hassan II de Casablanca;$^{(b)}$Facult\'{e} des Sciences, Universit\'{e} Ibn-Tofail, K\'{e}nitra;$^{(c)}$Facult\'e des Sciences Semlalia, Universit\'e Cadi Ayyad, LPHEA-Marrakech;$^{(d)}$LPMR, Facult\'e des Sciences, Universit\'e Mohamed Premier, Oujda;$^{(e)}$Facult\'e des sciences, Universit\'e Mohammed V, Rabat;$^{(f)}$Institute of Applied Physics, Mohammed VI Polytechnic University, Ben Guerir; Morocco.\\
% ...
$^{44}$Physics Department, Southern Methodist University, Dallas TX; United States of America.\\
% ...
$^{48}$Fakult\"{a}t Physik, Technische Universit{\"a}t Dortmund, Dortmund; Germany.\\
% ...
$^{54}$II. Physikalisches Institut, Georg-August-Universit\"{a}t G\"ottingen, G\"ottingen; Germany.\\
% ...
$^{83}$$^{(a)}$Khalifa University of Science and Technology, Abu Dhabi;$^{(b)}$New York University Abu Dhabi, Abu Dhabi;$^{(c)}$United Arab Emirates University, Al Ain;$^{(d)}$University of Sharjah, Sharjah; United Arab Emirates.\\
% ...
$^{102}$CPPM, Aix-Marseille Universit\'e, CNRS/IN2P3, Marseille; France.\\
% ...
$^{121}$Homer L. Dodge Department of Physics and Astronomy, University of Oklahoma, Norman OK; United States of America.\\
% ...
$^{154}$Raymond and Beverly Sackler School of Physics and Astronomy, Tel Aviv University, Tel Aviv; Israel.\\
% ...
$^{a}$ Also at Affiliated with an institute formerly covered by a cooperation agreement with CERN.\\
$^{b}$ Also at An-Najah National University, Nablus; Palestine.\\
$^{c}$ Also at Borough of Manhattan Community College, City University of New York, New York NY; United States of America.\\
$^{d}$ Also at Center for Interdisciplinary Research and Innovation (CIRI-AUTH), Thessaloniki; Greece.\\
$^{e}$ Also at Centre of Physics of the Universities of Minho and Porto (CF-UM-UP); Portugal.\\
$^{f}$ Also at CERN, Geneva; Switzerland.\\
% ...

\end{flushleft}
